\documentclass[preprint,showpacs,preprintnumbers,eqsecnum,amsmath,amssymb]{revtex4}
\usepackage{epsf}
\usepackage{color}
\usepackage{dcolumn}
\usepackage{bm}
\everymath{\displaystyle}
\newcommand{\sgn}{\operatorname{sgn}}

\begin{document} 

\title{Spin effects and compactification}

\author{Alexander J. Silenko}
\email{alsilenko@mail.ru} \affiliation{Research Institute for Nuclear Problems, Belarusian State University, Minsk 220030, Belarus\\ and
Bogoliubov Laboratory of Theoretical Physics, Joint Institute for Nuclear Research,
Dubna 141980, Russia}

\author{Oleg V. Teryaev}
\email{teryaev@theor.jinr.ru} \affiliation{Bogoliubov Laboratory
of Theoretical Physics, Joint Institute for Nuclear Research,
Dubna 141980, Russia}


\begin {abstract}
We consider the dynamics of Dirac particles moving in the curved spaces with one coordinate subjected
to compactification and thus interpolating smoothly between three- and two-dimensional spaces. We use
the model of compactification, which allows us to perform the exact Foldy-Wouthuysen transformation of
the Dirac equation and then to obtain the exact solutions of the equations of motion for momentum and spin
in the classical limit. The spin precesses with the variable angular velocity, and a ``flick'' may appear in the
remnant two-dimensional space once or twice during the period. We note an irreversibility in the particle
dynamics because the particle can always penetrate from the lower-dimensional region to the higher-dimensional
region, but not inversely. 
\end{abstract}

\maketitle

\section{Introduction}
Low-dimensional structures are now under scrutiny in
nonperturbative QCD, cosmology, high-energy physics,
and condensed matter physics. Properties of particles
placed into such structures are usually described by
considering quantum theory in two dimensions. However,
there is no doubt that real space remains three
dimensional, which may lead to qualitative differences in
some observables.

This especially concerns the particle spin properties,
which are crucially different at two and three spatial
dimensions (see, e.g., Refs.
\cite{Yip,DiracRep,DiracRe}). Thus, transition to (2+1)-dimensional spacetimes leads to losses of a significant part of such properties.
At the same time, in the two-dimensional space, anyons \cite{Chen:1989xs}
may appear.

In the present work, we investigate the problem of transformation of the spin properties under the compactification of some spatial dimension.
This problem is generally very difficult because the spin dynamics depends on many factors.
To extract some common properties, we consider the toy model
\cite{Fiziev1} of the curved space of \emph{variable} dimensionality 
smoothly changing from three to two. A great preference of the model used is a possibility to obtain \emph{exact} quantum-mechanical solutions.

We use the conventional Dirac equation for a consistent description of spin-1/2
particle motion in the curved space and take into
account relativistic effects. While such effects are not too important in condensed
matter physics (except for graphene), we keep in mind their further applications to the processes at
Large Hadron Collider in the case \cite{Mureika:LHC,StojkovicReview} of variable
(momentum) space dimension. We use the \emph{relativistic} method \cite{JMP} of
the Foldy-Wouthuysen (FW) transformation \cite{FW} to derive exact quantum-mechanical equations of motion 
and obtain their classical limit.

In this work, we focus our attention on the spin properties.
We show that, in contrast to a ``naive'' estimation, the
spin in an effectively two-dimensional space may precess
about the noncompactified dimensions and therefore a
``flick'' may appear in the remnant space once or twice
during the period. 

\section{Hermitian Hamiltonians
for the metric admitting the effective dimensional
reduction}\label{Hamiltonian}
%

Let us start with the following metric proposed by Fiziev \cite{Fiziev1}:
\begin{equation}
ds^2=c^2dt^2-\rho_1(z)^2d\Phi_1^2-\rho_2(z)^2d\Phi_2^2-\rho_3(z)^2dz^2,
\label{gmetric}
\end{equation}
where $\rho_3(z)^2=1+\rho'_1(z)^2+\rho'_2(z)^2$, the primes define derivatives
with respect to $z$, and $\rho_i$ are the functions of $z$.
The spatial coordinates vary in the limits $-\infty < z < \infty, ~ 0 < \Phi_{1,2} < 2 \pi$.

We suppose $\rho_i(z)$ to be positive.
The (3+1)-dimensional manifold defining this metric
is a hypersurface in a flat pseudo-Euclidean (5+1)-dimensional space.
The tetrad $e_{0}^{\widehat{0}}=1,~e_{i}^{\widehat{j}}=\delta^{ij}\sqrt{g_{ii}}$ allows us to define the local
Lorentz (tetrad) frame. This considerably simplifies an
analysis of results from possibly using the rescaled Cartesian coordinates $dX=\rho_1(z)d\Phi_{1}, ~dY=\rho_2 (z) d\Phi_{2},~dZ=\rho_3(z)dz$
in the neighborhood of any point.

Taking the limit $\rho_1(z) \rightarrow 0$ or the limit $\rho_2(z) \rightarrow 0$
may lead to the reduction of dimension of
the physical space from $d = 3$ to $d = 2$. We consider the case when the compactification
of the $\bm e_1~(\bm e_2)$ direction results in the confinement of the particle in
a narrow interval of $\Phi_1~(\Phi_2)$ angles.

The transverse part of the metric (if $z$ is assumed to be a
longitudinal coordinate) has the structure of the Clifford
torus, which is the product of two unit circles in the fourdimensional
Euclidean space: 
\begin{equation}
y_1^2+y_2^2=y_3^2+y_4^2=1.
\label{tmetric}
\end{equation}
The Clifford tori are used for analyzing twisted
materials \cite{clifford09} and vesicles \cite{cliffordva,cliffordve,cliffordvf}.
There is also some qualitative
similarity to projection of a tube in a six-dimensional space
onto a three-dimensional space, which was used for the
construction of the quasicrystals theory
\cite{Kalugin}.

We consider Clifford tori as a toy model of dimensional reduction.
We are not necessarily assigning the physical sense to all of the
intermediate values of $z$ except the asymptotics for $z \to\pm \infty$
corresponding to the three- and
two-dimensional spaces. Here, varying the dimension plays
the same role as varying the coupling constant for the case
of an adiabatic switch on the interaction.


To describe the spin-1/2 particles, we use the conventional covariant Dirac equation (see Ref. \cite{OSTgrav} and references therein).
To find the
Hamiltonian form of this equation,
one can substitute the given metric into the general equation
for the Hermitian Dirac
Hamiltonian (Eq. (2.21) in Ref. \cite{OSTRONG}). For the metric (\ref{gmetric}), the Hermitian Dirac
Hamiltonian was first derived in Ref. \cite{GorNeznArXiv}. It can be presented in the form
\begin{eqnarray}
{\cal H}_D =\beta mc^2  - \frac {i\hbar c}{ \rho_1}\alpha_1
\frac{\partial}{\partial\Phi_1}- \frac{ i\hbar c}{ \rho_2}\alpha_2
\frac{\partial}{\partial\Phi_2}- \frac {i\hbar c}{ 2}\alpha_3\left\{{\frac
1 \rho_3}, \frac{\partial}{\partial z}\right\},
\label{Hamilton2}\end{eqnarray} where $\{\dots,\dots\}$ denotes an
anticommutator.

We transform this Hamiltonian to the FW representation by the method elaborated in Ref. \cite{JMP}
which was earlier applied in our previous works
\cite{OSTgrav,OSTRONG,OST}.
After the {\it exact} FW transformation, we get the result
\begin{eqnarray}
{\cal H}_{FW} =\beta
\sqrt{a+{\hbar}\bm\Sigma\cdot\bm b},
\label{HamiltonFW}\end{eqnarray}
where
\begin{eqnarray}
a=m^2c^4+\frac{c^2p_1^2}{\rho^2_1}+\frac{c^2p_2^2}{\rho^2_2} + \frac{c^2}{4}
\left\{{\frac 1 \rho_3},
p_3\right\}^2,~~~ \bm{b}=b_1\bm e_1+b_2\bm e_2=\frac {c^2\rho'_2}{\rho^2_2\rho_3}p_2\bm e_1-\frac
{c^2\rho'_1}{\rho^2_1\rho_3}p_1\bm e_2,
\label{denot}\end{eqnarray}
and $(p_1,p_2,p_3)
=\left(-i\hbar\frac{\partial}{\partial\Phi_1},
-i\hbar\frac{\partial}{\partial\Phi_2},-i\hbar\frac{\partial}{\partial
z}\right)$ is the generalized momentum operator. Primes denote derivatives with respect to $z$.
The $\bm e_1,\bm e_2,\bm e_3$ vectors form the spatial part of the orthonormal basis
defining the local Lorentz (tetrad) frame.
For the given time-independent metric,
the operators ${\cal H}_{FW},~p_1,$ and $p_2$ are
integrals of motion.

Neglecting a noncommutativity of the $a$ and $\bm b$ operators allows us to omit anticommutators and results in
\begin{eqnarray}
{\cal H}_{FW} =\frac \beta
2\left(\sqrt{a+\hbar b}+\sqrt{a-\hbar b}\right)+\frac{\bm\Pi\cdot \bm
b}{2b}\left(\sqrt{a+\hbar b}-\sqrt{a-\hbar b}\right),
\label{HFW}\end{eqnarray} where $\bm\Pi=\beta\bm\Sigma$ is the spin polarization operator. It can be proven that extra terms appearing from the above noncommutativity are of order of
$|\hbar/(p_zl)|^3$, where $p_z$ is the particle momentum and $l$ is the characteristic
size of the nonuniformity region of the external
field (in the $z$ direction). With this accuracy,
\begin{eqnarray}
{\cal H}_{FW} =\beta \left(\sqrt{a}-\frac{\hbar^2b^2}{8a^{3/2}}\right)+\hbar\frac{\bm\Pi\cdot\bm b}{2\sqrt{a}}.
\label{HFWb}\end{eqnarray}
The second term proportional to $\hbar^2$ is important even when it is relatively small. This term contributes to the difference between gravitational interactions of spinning and spinless particles and therefore violates the weak equivalence principle. Its importance relative to the main term is defined by the ratio $(\hbar b/a)^2$. The weak equivalence principle is also violated by the spin-dependent Mathisson force (see Refs. \cite{OSTgrav,Plyatsko:1997gs} and references therein) defined by the third term in Eq. (\ref{HFWb}). While the third term is usually much bigger than the second one, it vanishes for unpolarized spinning particles. The second term proportional to $(\bm\Pi\cdot\bm b)^2$ is always nonzero. An analysis of Eqs. (\ref{denot}) and (\ref{HFWb}) leads to the conclusion that this term can be comparable with the main one (proportional to $\sqrt{a}$) when $l\sim\lambda_B$, where $\lambda_B$ is the de Broglie wavelength.
The existence of the term proportional to $\hbar^2$ is not a specific property of the toy model used. The appearance of such terms in the FW Hamiltonians describing a Dirac particle in Riemannian spacetimes was noticed in several works \cite{OST,DH,Jentschura},
whereas its relation to the spin-originated effect leading to the violation of the weak equivalence principle was never mentioned.

The equation of spin motion is given by
\begin{eqnarray}
{\frac {d{\bm \Pi}}{dt}} = \bm \Omega\times{\bm \Pi}, ~~~ \bm \Omega=
\beta\frac {\bm b}{\sqrt{a}}.
\label{finalOmegase}
\end{eqnarray}

As a result, the spin rotates
\emph{relative to} $\bm e_i$ \emph{vectors} $(i=1,2,3)$ with the angular velocity
$\bm \Omega$. Its motion relative to the Cartesian axes is much more complicated.

It has been proven in Ref. \cite{JINRLet1} that finding a classical limit of \emph{relativistic}
quantum mechanical equations reduces to the replacement of operators by
respective classical quantities when the condition of the Wentzel-Kramers-Brillouin approximation,
$\hbar/|pl|\ll1$, is satisfied.
It has also been shown that the classical limit of the FW Hamiltonians for Dirac \cite{OSTgrav,OSTRONG,OST} and scalar
\cite{Honnefscalar} particles in Riemannian spacetimes coincides with the corresponding purely
classical Hamiltonians.

\section{Motion of particle at variable dimensions}\label{Region}

Let us first study the motion of the particle by neglecting the influence of the spin onto its trajectory.
Since $p_1$ and $p_2$ are integrals of motion,
they can be replaced with the eigenvalues $\mathcal{P}_1$ and $\mathcal{P}_2$, respectively.
Let us  choose the $\bm
e_1$ axis as the compactified dimension and suppose that $\rho_{1}(z)$ is a decreasing function ($\rho_{1}(z)\rightarrow0$ when $z\rightarrow\infty$).
We can neglect a dependence of $\rho_{2}$ on $z$, assuming that this function changes much more slowly.
We denote initial values of all parameters by additional zero indices and consider the general case
when the initial value of the metric component, $\rho_{10}\equiv\rho_{1}(z_0)$, is
not small.

 The classical limit of the Hamiltonian is given by
\begin{eqnarray}
{\cal H}
=\sqrt{m^2c^4+\frac{c^2\mathcal{P}_1^2}{\rho^2_1}+\frac{c^2\mathcal{P}_2^2}{\rho^2_2}+\frac{c^2p_3^2}{\rho^2_3}}.
\label{HamiltonC}\end{eqnarray}


The possibility of making general conclusions with the special model used is based on the fact that the
Hamiltonian of a particle in an arbitrary static spacetime is given by
\begin{eqnarray}
{\cal H}
=\sqrt{\frac{c^2\left(m^2c^2+g^{ij}p_ip_j\right)}{g^{00}}}, ~~~ i,j=1,2,3.
\label{Hamiltong}\end{eqnarray}
Equation (\ref{Hamiltong}) covers spinless \cite{Cogn} and spinning \cite{OSTRONG,OSTgrav} particles in classical gravity as well as the classical limit of the corresponding quantum-mechanical Hamiltonians for scalar \cite{Honnefscalar}
and Dirac \cite{OSTgrav} particles. For spinning particles, the term  $\bm s\cdot\bm\Omega$ should be added to this Hamiltonian \cite{OSTRONG,OSTgrav}. When the metric is diagonal, $g^{ii}=1/{g_{ii}}$ and Eq. (\ref{Hamiltong}) takes the same form as Eq. (\ref{HamiltonC}).


To describe the compactification, we can introduce the compactification radius $\delta$ so that the
``compactification point'' $z_c$ can be defined by $\rho_1(z_c)=\delta$. Due to the energy $E$ conservation, the
particle can reach this point if
\begin{eqnarray}
E
\geq  \sqrt{m^2c^4+\frac{c^2\mathcal{P}_1^2}{\delta^2}+\frac{c^2\mathcal{P}_2^2}{\rho^2_2(z_c)}}.
\label{HamiltonE}\end{eqnarray}
Note that the decrease of compatification radius $\delta$ while $E$ remains finite implies the corresponding decrease of $\mathcal P_1$.

The particle velocity is equal to
\begin{equation}
v_z\equiv\frac {dz}{dt} =\frac{\partial{\cal H}}{\partial p_3}=
\frac{c^2p_3}{E\rho^2_3}\\=
c \frac{\sgn{(p_3)}}{E\rho_3(z)}\sqrt{E^2-m^2c^4- c^2 R(z)}, ~~~ R(z)=\frac{\mathcal{P}_1^2}{\rho^2_1(z)} + \frac{\mathcal{P}_2^2}{\rho^2_2(z)}.
\label{finalmw}
\end{equation}
Different signs correspond to the two different directions of the
longitudinal particle motion.
Note that the arrival to the compactification point with zero velocity ($z_c =z_f$ being the final point of particle trajectory) corresponds to
the equality sign in Eq. (\ref{HamiltonE}).

A tedious but simple calculation allows us to obtain the longitudinal component of the particle acceleration:
\begin{eqnarray}
a_z\equiv\frac {d^2z}{dt^2} =
- \frac{c^4}{E^2 \rho_3^2}\left(
\frac{R'}{2}+\frac{p_3^2\rho'_3}{\rho^3_3}
\right).
\label{vd}
\end{eqnarray}
It is obvious that  $p_3(z_f)=0,~ R'(z_f) \geq 0$ (for monotonic continuously differentiable $R(z)$), so that $a_z(z_f) \leq 0$. Therefore, $z_f$ is the turning (if $R'(z_f) > 0$) or attracting
(if $R'(z_f) = 0$) point. For nonmonotonic $R(z)$ there is a possibility of passage to the region $z > z_f$ due to possible growth of $\rho_2(z)$. The particle motion is then limited by the point
$\tilde z_f$ corresponding to the neglect of the motion in the $\bm e_2$ direction
\begin{eqnarray}
E
=  \sqrt{m^2c^4+\frac{c^2\mathcal{P}_1^2}{\rho_1(\tilde z_f)}}.
\label{tilde}\end{eqnarray}


The important particular case of Eq. (\ref{HamiltonC}) corresponds to $\mathcal{P}_1=0$. The particle penetrates into the region of the effective dimensional reduction ($z\rightarrow\infty$)
and does not reverse the direction of its motion.

In this study, as was mentioned above, we consider that
the smooth adiabatic transition from the three-dimensional
space to the effectively two-dimensional one does not
necessarily attribute the physical sense to all intermediate
points in particle motion. At the same time, the true change
of the dimensionality was discussed in cosmology (see
Refs. \cite{Mureika:2011bv,StojkovicReview,Fiziev:2010je,Sotiriou:2011xy})
and in connection with experiments at the LHC
(see Refs. \cite{StojkovicReview,Mureika:LHC,SCarlip,Sotiriou:LHC}). Our analysis 
can also be applicable at the LHC.

Note also that the motion in the opposite direction of increasing dimension does not impose any conditions for the
initial state of the particle. One may say that the region of lower dimension is ``repulsive'' whereas the region of higher dimension is ``attractive'',
implying a sort of \emph{irreversibility} in the particle dynamics. This property emerges because of the appearance of $\rho_1$
in the expression for the Hamiltonian in the denominator. Such a situation is a general one that can be seen from Eq. (\ref{Hamiltong}) in the case of diagonal metric.
This may give additional support to the hypothesis \cite{Mureika:2011bv,Fiziev:2010je}
that such a transition from the lower dimensionality to the higher one leaded to the evolution
of the Universe.


\section{Spin evolution at variable dimensions}

In the classical limit, the angular velocity of spin precession is given by
\begin{eqnarray}
\bm \Omega=
\frac {\bm b}{E}=\frac{c^2}{E\rho_3}\left(
\frac{\mathcal{P}_2\rho'_2}{\rho^2_2}\bm e_1-\frac
{\mathcal{P}_1\rho'_1}{\rho^2_1}\bm e_2\right).
\label{finalOmegacl}
\end{eqnarray}
Because $d\bm s/dt=v_z(z)(d\bm s/dz)$, Eqs. (\ref{finalmw}) and (\ref{finalOmegacl})
define an easily solvable system of first-order homogeneous linear differential equations.

Equation (\ref{finalOmegacl}) is rather informative about details of the compactification. Only the $\Omega_2$ component contains parameters of the compactified dimension.
Although $|\mathcal{P}_1|/|\mathcal{P}_2|\ll1$, the presence of additional factors
does not allow for 
neglecting $\Omega_2$ as compared with $\Omega_1$ (under the condition that $\mathcal{P}_1\neq0$).

When
$\rho_2(z)=const$, $\Omega_1=0$ and the spin rotates about the
$\bm e_2$ axis, the spin projection onto the $\bm e_2\bm e_3$ surface, which
is the spatial part of the (2+1)-dimensional spacetime,
oscillates. The spin appears in this surface only once (in the
special case when the cone of spin precession is tangent to
this surface) or twice per rotation period. Evidently, the
origin of this spin ``flickering'', as well as the appearance of
pseudovector, is completely unexplainable in terms of the
two-dimensional space.


The model used allows to obtain an exact analytical description of the spin evolution.
It is characterized by a change of the angle $\varphi$ defining the direction
of the spin in the plane orthogonal to $\bm\Omega$:
\begin{eqnarray}
 \Delta\varphi(z) =\int{\Omega(t)dt}=\int_{z_{0}}^{z}{\frac{\Omega(y)}{v_z(y)}dy}.
\label{varph}
\end{eqnarray}

The problem of spin evolution at the effective dimensional reduction can be solved in a general form.
To simplify the
analysis, let us consider the case of $\rho_2(z)=\rho_{20}=const$.
In this case, the \emph{exact} value of the integral is
\begin{eqnarray}
\Delta\varphi(z)=
\arcsin{\frac{c\mathcal{P}_1}{A\rho_1(z)}}-
\arcsin{\frac{c\mathcal{P}_1}{A\rho_{10}}},
\label{varin}
\end{eqnarray}
where
\begin{eqnarray} A=\sqrt{E^2-m^2c^4-\frac{c^2\mathcal{P}_2^2}{\rho^2_{20}}}=
c \sqrt{\frac{p_{30}^2}{\rho^2_{30}}+\frac{\mathcal{P}_1^2}{\rho^2_{10}}}.
\label{eqna}
\end{eqnarray}
Since
\begin{eqnarray}
\rho_{1}(z_f)=c\left|\mathcal{P}_{1}\right|\left(E^2-m^2c^4-\frac{c^2\mathcal{P}_2^2}{\rho^2_{20}}\right)^{-1/2},
\label{zetf}
\end{eqnarray}
the total spin turn ($z=z_f$) 
is given by
\begin{eqnarray}
\Delta\varphi=\sgn{(\mathcal{P}_1)}\cdot\frac\pi 2-
\arctan{\frac{\mathcal{P}_1\rho_{30} }{\rho_{10}p_{30}}}.
\label{varif}
\end{eqnarray}
The passage of the particle to the region of compactification implies, as was discussed above,
the relative smallness of the second term so that
the spin rotates by about $90^\circ$.

If $\mathcal{P}_1=0$, 
the spin projection onto the $\bm e_1$ direction is always conserved. The spin can, however, rotate about the $\bm e_1$ direction if $\rho_2$ depends on $z$. In this case, the angle of the spin turn is equal to
\begin{eqnarray}
\Delta\phi(z)=-\arcsin{\frac{c\mathcal{P}_2}{B\rho_2(z)}}+
\arcsin{\frac{c\mathcal{P}_2}{B\rho_{20}}}, ~~~ 
B=\sqrt{E^2-m^2c^4}=
\sqrt{\frac{c^2p_{30}^2}{\rho^2_{30}}+\frac{c^2\mathcal{P}_2^2}{\rho^2_{20}}}.
\label{varfi}
\end{eqnarray}

The total spin turn ($z=z_f$) is given by
\begin{eqnarray}
\Delta\phi=
\arctan{\frac{\mathcal{P}_2\rho_{30} }{\rho_{20}p_{30}}}-\sgn{(\mathcal{P}_2)}\cdot\frac\pi 2.
\label{varff}
\end{eqnarray}

\section{Conclusions and outlook}
We considered the Dirac fermion dynamics in the curved
space model of variable dimension. The advantage of the
toy model used is the possibility of performing the exact
FW transformation of the Dirac equation and then
obtaining the exact solutions of the equations of motion
for momentum and spin in the classical limit. At the same
time, the obtained Hamiltonian (\ref{HamiltonC}) is  similar to the generic one  (\ref{Hamiltong}) so that one can expect that
qualitative features of spin and momentum dynamics will persist for other compactification-related metrics as well.

The analysis of particle momentum evolution allows us to describe the motion at the boundary between the regions of space
having different dimensions.
The passage to the region of lower dimension is more natural in the special case when the generalized momentum in the compactified direction $\mathcal{P}_1=0$.
At the same time, the transition to the region of higher dimension (considered in Refs. \cite{Mureika:2011bv,Fiziev:2010je} as a possible way of the evolution of the Universe) does not impose the constraints for its initial state, manifesting a sort of irreversibility.

The particle motion (especially near the turning point) is characterized by the three main properties which cannot be naturally explained from the point of view of
observer residing in the compactified spacetime:
\emph{i)} a reversion of the direction of motion; \emph{ii)} a rather quick motion along the compactified direction, which may be seen as a sort of ``zitterbewegung'';
\emph{iii)} the appearance of a pseudovector of spin in the compactified  (2+1)-dimensional space and its
rotation or flickering [when the spin pseudovector crosses
the remnant (2+1)-dimensional layer]. 

The experimental tests of the emerging spin effects may be performed by studies of spin polarizations of $\Lambda$ (and, probably, also $\Lambda_c$) hyperons produced in the high-energy collisions where the compactification \cite{Mureika:LHC,StojkovicReview}
takes place. This may bear a resemblance to the recently proposed \cite{Baznat:2013zx}
tests of the vorticity in heavy-ion collisions, although a detailed analysis is required.

We can finally conclude that the transition to (2+1)-dimensional spacetime 
leads to the nontrivial behavior of spin which, generally speaking, cannot be adequately described from the point of view of an observer residing at (2+1) dimensions.

\section*{Acknowledgments}

We are indebted to P.P. Fiziev,  V.P. Neznamov, and D.V. Shirkov for stimulating discussions.
This work was supported in part by the RFBR (Grants No. 11-02-01538  and 12-02-91526) and BRFFR
(Grant No. $\Phi$12D-002).


\end{document}